\documentstyle[12pt]{article}
\begin{document}
\begin{titlepage}
\author{E. Dudas \\ \\ CEA, Service de Physique Th\'eorique, CE-Saclay \\
F-91191 Gif-sur-Yvette Cedex, FRANCE}
\title{\bf Holomorphy and dynamical scales in supersymmetric gauge theories}
\maketitle
\vskip 1cm
\begin{abstract}
New recent results in supersymmetric gauge theories based on holomorphy and
symmetry considerations are extended to the case where the gauge coupling
constant is given by the real part of a chiral superfield. We assume here
that its dynamics can be described by an effective quantum field theory.
Then its vacuum expectation value is a function of the other coupling
constants, viewed as chiral background superfields. This functional
dependence can be determined exactly and satisfies highly non-trivial
consistency checks.
\end{abstract}
\vskip 2cm
\begin{flushright}
Saclay T95/123 \\ November 1995
\end{flushright}
\end{titlepage}
\newpage\

\section{Introduction}

In the last year, a lot of progress has been done in understanding the
dynamics of $N=1$ supersymmetric gauge field theories in four dimensions
\cite{1-intri}. Using the principle of holomorphy introduced in \cite
{seib318} (see also \cite{11-shifman}, \cite{10-amati}, \cite{14-kapluno}),
 new results were obtained for supersymmetric QCD in case where
the number of flavours $N_f$ is larger or equal to the number of colours
\cite{seib49},\cite{4-seib435}.\ New exact superpotentials were derived with
a highly non-trivial dynamics \cite{5-intri50} and with a possible impact on
supersymmetry breaking \cite{6-intri342}.

Some of the results were checked by dynamical instanton computations \cite
{7-finnell}.\ The methods were further generalized in order to accommodate
for soft-breaking terms \cite{8-evans}, \cite{9-aharony}.

The two main tools used can be described as follows.

\noindent (1) Holomorphy: The Wilsonian superpotential $W_{{\rm eff}}$ is a
holomorphic function of the fields and coupling constants.\ This is
equivalent to consider the coupling constants as chiral background fields.

\noindent (2) Symmetries: $W_{{\rm eff}}$ is invariant under all the
symmetries of the model.\ If some of the symmetries are explicitly broken by
the coupling constants, we assign transformation laws to the coupling
constants such as to restore the full symmetry group.\ The anomalous
symmetries can be treated on the same footing by assigning specific
transformation laws to the scale(s) $\Lambda $ of the gauge group(s)
\cite{5-intri50}.

In case where supersymmetry is not spontaneously broken, the combination of
these two principles implies that the vacuum expectation values of chiral
fields in $W_{{\rm eff}}$ are holomorphic functions of the coupling
constants compatible with the symmetries.\ This is a powerfull result which
will be largely used in the following.

The purpose of this letter is to apply these methods to the case where the
gauge coupling constant is promoted to a dynamical chiral superfield (the
dilaton superfield of effective superstring theories).\ If the
dynamics of this additional field can be described by an effective quantum
field theory, then by use of the above-mentioned principles, its vacuum
expectation value is a function of the other coupling constants of the
theory. This is a strong assumption; usually it is assumed that the gauge
coupling is determined by the high energy physics. The new hypothesis
gives new relations, and their consistency with the already known
dynamic relations is a highly non-trivial necessary consistency check.
Still, our considerations can only prove in which case this assumption
is wrong; it can well be that all our consistency checks are verified
and for other reasons, the gauge coupling is fixed by the high energy
dynamics.

Section 2 describes the consequences of this assumption in the case of
supersymmetric QCD with gauge group ${\rm SU}(N_c)$ and number of flavours $%
N_f\geq N_c$ (with or without an additional field in the adjoint of the
gauge group) and gauge group $SO(N_c)$. Some consistency checks are
worked out in detail, related to the decoupling of massive flavors or
of the adjoint field, the NSVZ beta function to all orders
\cite{12-novikov} and the non-abelian duality of ref. \cite{4-seib435}.

In section 3 an effective Lagrangian analysis is performed, which supports
and clarifies the results of section 2.

Finally some conclusions are drawn together with some comments.
\section{Dynamical gauge coupling constant and holomorphy}

The model to be considered below is a simple generalization of
supersymmetric QCD with $N_f$ flavours of quarks, $Q^i$ in the $N_c$
representation of ${\rm SU}(N_c)$ gauge group and $\widetilde{Q}_{{\bar{%
\imath}}}$ in the $\overline{N}_c$ representation. For simplicity
reasons, we begin with the case $N_F=N_c=N.$
The Lagrangian is given by

\[
{\cal L}={\cal L}_{{\rm Kin}}+{\cal L}_{{\rm couplings}} \ ,
\]

\noindent where

\begin{eqnarray}
{\cal L}_{{\rm Kin}} &=&\int d^4\theta \left[ Q^{+}{\rm e}^VQ+\widetilde{Q}\,%
{\rm e}^{-V}\widetilde{Q}^{+}+K\left( S^{+},S\right) \right] +  \label{un} \\
&&+\left( \int d^2\theta \frac 14SW^\alpha W_\alpha +h.c.\right)   \nonumber
\end{eqnarray}

\noindent and

\begin{equation}
{\cal L}_{{\rm couplings}}=\int d^2\theta \left( m_i^{{\bar{\imath}}} Q^i%
\widetilde{Q}_{{\bar{\imath}}}+bB+\widetilde{b}\widetilde{B}\right) +h.c. \ .
\label{deux}
\end{equation}

The Wilson gauge coupling constant is $\frac 1{g_W^2}=\left\langle {Re}
S\right\rangle ,$ where $S$ is a chiral superfield.\ In (\ref{deux}), we
introduced the composite chiral superfields (the ``baryons'')

\begin{equation}
B =\frac 1{N!}\epsilon _{i_1...i_N} \;Q^{i_1}...Q^{i_N}, \ \ \
\widetilde{B} =\frac 1{N!}\epsilon ^{{\bar{\imath}}_1...{\bar{\imath}}_N}\;%
\widetilde{Q}_{{\bar{\imath}}_1}...\widetilde{Q}_{{\bar{\imath}}_N} \ ,
\label{trois}
\end{equation}

\noindent $b,\widetilde{b}$ are associated sources and $m_i^{{%
\bar{\imath}}}$ is a quark mass matrix in the flavour space
\noindent $\left( i,{\bar{\imath}}=1,...,N_f\right) .$

The global symmetry of the model, in the sense described in the
Introduction, is

\begin{equation}
G={\rm SU}(N_f)\times {\rm SU}(N_f)\times U(1)_B\times U(1)_R\times
U^{\prime }(1)_R  \label{quatre}
\end{equation}

\noindent where ${\rm SU}(N_f)\times {\rm SU}(N_f)$ is the flavour chiral
symmetry and $U(1)_B$ the baryon conservation.\ The most important point in our
discussion are the two R-symmetries $U(1)_R\times U^{\prime }(1)_R.$ The $%
U(1)_R$ symmetry is always spontaneously broken and acts on the fields and
couplings as follows:

\begin{eqnarray}
\left( Q,\widetilde{Q}\right) \left( \theta ^{\prime }={\rm e}^{- \frac{%
3i\beta }2}\theta \right)  &=&{\rm e}^{-i\beta }\left( Q,\widetilde{Q}%
\right) \left( \theta \right) ,  \label{cinq} \\
W_\alpha ^{\prime }\left( \theta ^{\prime }\right)  &=&{\rm e}^{\frac{%
-3i\beta }2}W_\alpha \left( \theta \right) , \
S^{\prime }\left( \theta ^{\prime }\right)  =S\left( \theta \right) +\frac{%
ibo}{8\pi ^2}\beta ,  \nonumber \\
\left( m^{\prime }\right) _i^{\bar{\imath}} &=&{\rm e}^{-i\beta
}m_i^{\bar{\imath}}, \ \left( b,%
\widetilde{b}\right) ^{\prime }={\rm e}^{i\left( N-3\right) \beta} \left( b,%
\widetilde{b}\right) .  \nonumber
\end{eqnarray}

The non-linear transformation of the field S, with coefficient $%
b_0=3N_c-N_f=2N$ in our case, cancels the one-loop triangle anomalies and
restores the symmetry.

The $U^{\prime }(1)_R$ symmetry is the anomaly-free combination of the usual
axial symmetry and an R-symmetry. Its action is

\begin{eqnarray}
\left( Q,\widetilde{Q}\right) \left( \theta ^{\prime }={\rm e}^{i\gamma
}\theta \right)  &=&\left( Q,\widetilde{Q}\right) \left( \theta \right) ,
\label{six} \\
W_\alpha ^{\prime }\left( \theta ^{\prime }\right)  &=&{\rm e}^{i\gamma
}W_\alpha \left( \theta \right) ,\;S^{\prime }\left( \theta ^{\prime
}\right) =S\left( \theta \right) ,  \nonumber \\
\left( m^{\prime }\right) _i^{\bar{\imath}} &=&{\rm e}^{2i\gamma }m_i^
{\bar{\imath}},\;\left( b,%
\widetilde{b}\right) ^{\prime }={\rm e}^{2i\gamma }\left( b,\widetilde{b}%
\right) .  \nonumber
\end{eqnarray}

The low-energy effective theory is described by the gauge invariant fields $%
B,\widetilde{B}$ introduced in (\ref{trois}), the meson fields $M_{{\bar{%
\imath}}}^i=Q^i\widetilde{Q}_{{\bar{\imath}}}$ and the gauge coupling
superfield S (the glueball superfield ${\rm %
U}={\rm tr}W^\alpha W_\alpha $ is introduced also for convenience). Using
the principles (\ref{un}) and (\ref{deux}) in the Introduction, their vacuum
expectation values are given by

\begin{eqnarray}
\left\langle M_{{\bar{\imath}}}^i\right\rangle  &=&C_M\left( \frac{\det m}{b%
\widetilde{b}}\right) ^{\frac 1{N-2}}\left( \frac 1m\right) _{{\bar{\imath}}%
}^i\;,  \label{sept} \\
\left\langle B\right\rangle  &=&C_B\left( \frac{\det m}{b^{N-1}\widetilde{b}%
}\right) ^{\frac 1{N-2}}\;,\left\langle \widetilde{B}\right\rangle =C_{%
\widetilde{B}}\left( \frac{\det m}{b\widetilde{b}^{^{N-1}}}\right) ^{\frac
1{N-2}}\;,  \nonumber \\
\left\langle U\right\rangle  &=&C_U\cdot \left( \frac{\det m}{b\widetilde{b}}%
\right) ^{\frac 1{N-2}}\equiv \Lambda _{N,0}^3\;,  \nonumber \\
\left\langle S\right\rangle  &=&C_S-\frac 1{4\pi ^2(N-2)}\ln \frac{\det m}{%
\mu ^{N(N-2)}\left( b\widetilde{b}\right) ^{\frac N2}}%
\;.  \nonumber
\end{eqnarray}

In (\ref{sept}), $C_M...C_S$ are some numbers to be more constrained by
dynamical considerations, $\Lambda _{N,0}$ is the scale of the theory in
case where all the quarks are heavy and decouples and $\mu $ is a mass scale
introduced in order to restore the dimensions in $\left\langle
S\right\rangle $, consistently interpreted as a renormalization scale in
the following. The dynamical scale of the theory $\Lambda _{N,N}$ is
defined, as usual, by the 1-loop running of the Wilson gauge coupling
constant:

\begin{equation}
\Lambda _{N,N}=\mu \ {\rm e}^{-\frac{4\pi ^2\left\langle S\right\rangle }%
N}=C_\Lambda \cdot \left[ \frac{\det m}{{\left( b\widetilde{b}\right)}^
{N \over 2}} \right] ^{\frac 1{N(N-2)}},  \label{huit}
\end{equation}

\noindent where $C_\Lambda ={\rm e}^{-\frac{4\pi ^2C_S}N}.$
One remark concerning the comparison of $<M_{\bar{\imath}}^i>$, $<B>$,
$<\widetilde B>$ in (\ref{sept}) with the results of ref. \cite{seib49} is
 in order.
Using the result (\ref{huit}) of our additional assumption in eqs. (4.12)-
(4.15) of ref. \cite{seib49}, we find that the power series in eqs. (4.14)
 and (4.15)
in \cite{seib49} collapse to only one term and the final result is
 qualitatively of
the form (\ref{sept}). We will come back to this point in the next paragraph.
Combining (\ref{sept}) and (\ref{huit}) in a straightforward way we get

\begin{equation}
\Lambda _{N,0}=\frac{C_U^{1/3}}{C_\Lambda ^{2/3}}\left( \det m\right)
^{\frac 1{3N}}\Lambda _{N,N}^{2/3}=\frac{C_U^{1/3}}{C_\Lambda ^{N/3}}\left( b%
\widetilde{b}\right) ^{1/6}\Lambda _{N,N}^{\frac N3}\;.  \label{neuf}
\end{equation}

The eqs. (\ref{neuf}) express the well-known relations between $\Lambda
_{N,0}$ and $\Lambda _{N,N}$ in two particular cases. Namely, if $b,%
\widetilde{b}\rightarrow 0$ a direct one-loop computation gives $\Lambda
_{N,0}=\left( \det m\right) ^{\frac 1{3N}}\Lambda _{N,N}^{\frac 23}.$ If $%
m\rightarrow 0,$ a similar computation gives $\Lambda _{N,0}=\left( b%
\widetilde{b}\right) ^{\frac 16}\Lambda _{N,N}$ \cite{seib49}. Our
expression (\ref{huit}) actually encodes both limits in an exact
formula.\ The correct step function decoupling in (\ref{neuf}) fixes
the coefficients $C_U=C_\Lambda =1.$

An additional dynamical constraint comes from the Konishi anomaly (see \cite
{10-amati}), which in our case reads

\begin{equation}
\left\langle m_i^{{\bar{\imath}}}M_{{\bar{\imath}}}^i+bB\right\rangle
=-\left\langle U\right\rangle \;,\left\langle bB\right\rangle =\left\langle
\widetilde{b}\widetilde{B}\right\rangle \;.  \label{dix}
\end{equation}

\noindent This results in $C_B=C_{\widetilde{B}}$ and $C_M+C_B=-1.$ To
summarize, consistency of (\ref{sept}) with the decoupling theorem and the
Konishi anomaly (\ref{dix}) fix four of the five unknown coefficients in (%
\ref{sept}),

\begin{equation}
C_U=C_\Lambda =1\;,C_M+C_B=-1\;,C_B=C_{\widetilde{B}}\;.  \label{onze}
\end{equation}

It is interesting to note that using (\ref{huit}) and (\ref{onze}) into (\ref
{sept}) we can write

\begin{eqnarray}
\left\langle M_{\bar{\imath}}^i\right\rangle  &=&C_M\Lambda _{N,N}^2(\det
m)^{\frac
1N}\left( \frac 1m\right) _{{\bar{\imath}}}^i\;,  \label{douze} \\
\left\langle B\right\rangle  &=&C_B\Lambda _{N,N}^N\left( \frac{\widetilde{b}%
}b\right) ^{1/2}\;,\left\langle \widetilde{B}\right\rangle =C_B\Lambda
_{N,N}^N\left( \frac b{\widetilde{b}}\right) ^{1/2}\;.  \nonumber
\end{eqnarray}

These relations were usually obtained in some limits, the first in the $b,%
\widetilde{b}\rightarrow 0$ limit and the second line in the $m\rightarrow 0$
limit \cite{seib49}. Hence, considering the gauge coupling as a dynamical field
seems to somehow relate the two
limits where $m\rightarrow 0$ and $b,\widetilde{b}\rightarrow 0.$

We turn now to the case $N_f>N_c$. The superpotential in this case reads

\begin{equation}
W_{{\rm couplings}}=m_i^{\bar{\imath}}Q^i\widetilde{Q}_{\bar{\imath}}+
\sum_{a} m_aQ^{N_c +a}\widetilde{Q}_{N_c +a}+bB+%
\widetilde{b}\widetilde{B}\;,  \label{treize}
\end{equation}

\noindent where $i,{\bar{\imath}}=1,...,N_c\;,a=1,...,N_f-N_c \ $ and
$\left( B,\widetilde{B}\right) =\left( B^{N_c+1,...,N_f},\widetilde{B}%
_{N_c+1,...,N_f}\right) .$ We therefore added specific source terms in (\ref
{treize}) for simplicity reasons.\ The global symmetry group in this case is

\begin{eqnarray}
G &=&SU\left( N_f\right) \times SU\left( N_f\right) \times U\left( 1\right)
_B\times \left( \prod_aU\left( 1\right) _{Ba}\right) \times
\label{quatorze} \\
&&\times U\left( 1\right) _R\times U^{\prime }\left( 1\right) _R\times
\left( \prod_aU\left( 1\right) _{Ra}\right) .  \nonumber
\end{eqnarray}

\noindent In (\ref{quatorze}), $U\left( 1\right) _{Ba}$ and $U\left(
1\right) _{Ra}$ are the baryon number conservation of the heavy quarks and
R-symmetries to be displayed below.

The $U\left( 1\right) _R$ symmetry acts as in (\ref{cinq}) with the
appropriate beta function coefficient $b_0=3N_c-N_f.$ The action of $%
U' \left( 1\right) _R$ is

\begin{eqnarray}
\left( Q,\widetilde{Q}\right) \left( \theta ^{\prime }={\rm e}^{i\gamma
}\theta \right)  &=&{\rm e}^{\frac{i\left( N_f-N_c\right) \gamma }{N_f}%
}\left( Q,\widetilde{Q}\right) \left( \theta \right) ,  \label{quinze} \\
W_\alpha ^{\prime }\left( \theta ^{\prime }\right)  &=&{\rm e}^{i\gamma
}W_\alpha \left( \theta \right) \;,S^{\prime }\left( \theta ^{\prime
}\right) =S\left( \theta \right) \;,  \nonumber \\
m^{\prime } &=&{\rm e}^{\frac{2iN_c\gamma }{N_f}}m \ ,\;\left( b,\widetilde{b}%
\right) ^{\prime }={\rm e}^{\frac{i\left( N_c^{2}-N_cN_f+2N_f\right)
\gamma }{N_f}}\left( b,\widetilde{b}\right) \;,  \nonumber
\end{eqnarray}

\noindent where $Q,\widetilde{Q} \ (m)$ are the set of all quark fields
(masses) of the theory. The action of $U(1)_{Ra}$ is given by

\begin{eqnarray}
\left( Q^i,\widetilde{Q}_{\bar{\imath}}\right) \left( \theta ^{\prime }={\rm
e}^{i\delta
}\theta \right)  &=&{\rm e}^{i\delta }\left( Q^{i},\widetilde{Q}%
_{\bar{\imath}}\right) \left( \theta \right) \;,  \label{seize} \\
\left( Q^{N_c+b},\widetilde{Q}_{N_c+b}\right) \left( \theta ^{\prime
}\right)  &=&{\rm e}^{-i\left( N_c\delta _{ab}-1\right) \delta }\left(
Q^{N_c+b},\widetilde{Q}_{N_c+b}\right) \left( \theta \right) \;,  \nonumber
\\
W_\alpha ^{\prime }\left( \theta ^{\prime }\right)  &=&{\rm e}^{i\delta
}W_\alpha \left( \theta \right) \;,S^{\prime }\left( \theta ^{\prime
}\right) =S\left( \theta \right) \;,  \nonumber \\
\left( m_i^{\bar{\imath}}\right) ^{\prime } &=&m_i^{\bar{\imath}}\;,m^{\prime
}_b={\rm e}^{2iN_c\delta
_{ab}}m_b\;,  \nonumber \\
\left( b,\widetilde{b}\right) ^{\prime } &=&{\rm e}^{i \left( 2-N_c\right)
\delta }\left( b,\widetilde{b}\right) \;.  \nonumber
\end{eqnarray}

\noindent Using the same strategy as before, we get

\begin{equation}
\left\langle S\right\rangle =-\frac 4{N_c-2}\ln \frac{\det m_i^{{\bar{\imath}%
}}}{\mu^{\frac{\left( N_c-2\right) \left( 3N_c-N_f\right) }2} \left( b%
\widetilde{b}\right) ^{\frac{N_c}2}\prod\limits_am_a^{\frac{N_c-2}2}}\;.
\label{dix-sept}
\end{equation}

\noindent Defining the dynamical scale $\Lambda _{N_c,N_f}=\mu \cdot {\rm e}%
^{-\frac{8\pi ^2\left\langle S\right\rangle }{3N_c-N_f}}$ and comparing with
$\Lambda _{N_c,N_c}$ in $(\ref{huit})$, we get the relation

\begin{equation}
\Lambda _{N_c,N_c}=\left( \prod\limits_am_a^{\frac 1{2N_c}}\right) \cdot
\Lambda _{N_c,N_f}^{\frac{3N_c-N_f}{2N_c}}\;.  \label{dix-huit}
\end{equation}

\noindent The eq. (\ref{dix-huit}) is exactly that we would get in a
one-loop computation with threshold effects at the scales $m_a.$ One can
also check that in this case

\begin{equation}
\left\langle U\right\rangle =\left( \frac{\det m_i^{{\bar{\imath}}}}{b%
\widetilde{b}}\right) ^{\frac 1{N_c-2}}\equiv \Lambda _{N_c,0}^3
\label{dix-neuf}
\end{equation}

\noindent and we get the well known equation$\;\Lambda _{N_c,0}=\left( \det
m_i^{\bar{\imath}}\cdot \prod\limits_am_a\right) ^{\frac 1{3N_c}}\cdot \Lambda
_{N_c,N_f}^{\frac{3N_c-N_f}{3N_c}}\;.$ It is interesting to examine eq.(\ref
{dix-sept}) closer in connection with the measurable gauge coupling constant
$g.$ It is known that $\Lambda _{N_c,N_f}$, $g$ and $g_W$ are
related through \cite{11-shifman},\cite{12-novikov}

\begin{equation}
\Lambda _{N_c,N_f}=\mu \ {\rm e}^{-\frac{8\pi ^2}{\left( 3N_c-N_f\right)
g_W^2}}=\frac \mu {g^{\frac{2N_c}{3N_c-N_f}}} \ {\rm e}^{-\frac{8\pi ^2}{%
\left( 3N_c-N_f\right) g^2}}\;.  \label{vingt}
\end{equation}

\noindent Combining (\ref{dix-sept}) and (\ref{vingt}) we find the exact
renormalization group evolution of $g$

\begin{equation}
\frac 1{g^2\left( \mu \right) }=\frac 1{g^2\left( M_X\right) }-\frac{\left(
3N_c-N_f\right) }{8\pi ^2}\ln \left[ \left( \frac{g(\mu )}{g(M_X)}\right) ^{%
\frac{2N_c}{3N_c-N_f}} \ Z_Q^{\frac{N_f}{3N_c-N_f}} \ \frac{M_X}\mu
\right] \;.  \label{vingt-un}
\end{equation}

\noindent In (\ref{vingt-un}), $M_X$ is a high scale and $Z_Q$ is defined by
$m\left( \mu \right) =Z_Q^{-1}\;m(M_X),b\left( \mu \right) =Z_Q^{-\frac{N_c}%
2}b(M_X),$ where we denote by $m$ the set of the masses $m_i^{{\bar{\imath}}%
},m_a.$ Differentiating (\ref{vingt-un}) with respect to $\mu $ and defining
the anomalous dimension $\gamma _Q=-\frac{\partial \ln Z_Q}{\partial \ln \mu }
$, we find the exact beta function

\begin{equation}
\beta (g)=-\frac{g^3}{16\pi ^2} \ \frac{3N_c-N_f+N_f\;\gamma_Q}{1-\frac{%
N_c}{8\pi ^2}g^2}\;.  \label{vingt-deux}
\end{equation}

\noindent The expression (\ref{vingt-deux}) is known since a long time \cite
{12-novikov}, \cite{11-shifman} to be the beta function in all orders in
supersymmetric $QCD.$ We consider this as an independent consistency check
of our relation (\ref{dix-sept}).

The one-loop running of $g_W$ is easily understood taking into account that
the holomorphic parameters $m,b,\widetilde{b}$ entering into eq.(\ref
{dix-sept}) are not physical, being related to the physical ones by the
renormalization constant $Z_Q$ defined above. By the
non-renormalization theorem, $m,b,\widetilde{b}$ do not run.
Hence one recovers the usual one loop running

\begin{equation}
\frac 1{g_W^2\left( \mu \right) }=\frac 1{g_W^2\left( M_X\right) }-\frac{%
3N_c-N_f}{8\pi ^2}\ln \frac{M_X}\mu \;.  \label{vingt-trois}
\end{equation}

Finally notice that the gaugino condensation scale $\left\langle
U\right\rangle =\Lambda _{N_c,0}^3$ is renormalization group invariant, as
it should be.

Notice that for $N_f < N_c$ the baryons and the sources $b, \widetilde{b}$
do not exist. In this case it is easy to check that there are more $R$
symmetries than chiral background parameters and it is impossible to
write down expressions similar to (\ref{huit}),
(\ref{dix-sept})-(\ref{dix-huit}). We therefore recover the well-known
runaway behaviour in SQCD for $N_f < N_c$.

These results are easily generalized to other models.\ We checked, for
example, the case of an additional chiral superfield $X$ in the adjoint of
the gauge group studied in \cite{15-kutasov}, or models with gauge groups $%
SO\left( N_c\right) $ studied in \cite{4-seib435},\cite{16-intri}.
In the first case, the superpotential of the model is ($k$ is fixed)

\begin{equation}
W_{{\rm couplings}}=g_k Tr X^{k+1} +
m_i^{\bar{\imath}}Q^i\widetilde{Q}_{\bar{\imath}}+
\sum_{a} m_aQ^{N_c +a}\widetilde{Q}_{N_c +a}+bB+%
\widetilde{b}\widetilde{B}\;.  \label{trente}
\end{equation}
The holomorphy and the symmetries give

\begin{equation}
\left[ {\Lambda_{N_c,N_f}^{(X)}}\right] ^{ (2N_c-N_f)} =
\left( \det m_i^{\bar{\imath}} \prod\limits_am_a\right)
^{-1} g_k^{-{2N_c \over k+1}} \left[ \frac{\det m_i^{{\bar{\imath}}}}{ b%
\widetilde{b}}\right] ^{\frac {2N_c}{\left( N_c-2\right)
\left( k+1\right) }}  \ . \label{trente et un}
\end{equation}

\noindent The gaugino condensation scale is the same as before,
eq.(\ref{dix-neuf}).
In the second case, we have $2N_f$ quark fields $Q^i$. Defining the baryons
$B$ as in (\ref{trois}) and considering the superpotential

\begin{equation}
W_{{\rm couplings}}={1 \over 2} m_{ij} Q^i Q^j+
\sum_{a} m_a Q^{N_c+a} Q^{N_c+a} +bB \;,  \label{trente-deux}
\end{equation}

\noindent we get

\begin{equation}
{\Lambda_{N_c,N_f}^{(SO(N_c))}} =
\left(b^2 \prod\limits_am_a\right)
^{-{1 \over 3N_c-N_f-6}}  \label{trente-trois}
\end{equation}

\noindent and the gaugino condensation scale is
 ${\Lambda_{N_c,0}^{(SO(N_c))}} =
\left( \frac {\det m_{ij}}{b^2} \right)  ^{\frac 1{3(N_c-2)}}$.

 In all
these cases, the expression of the dynamical scale as a function of the
background parameters gives a correct decoupling of heavy fields and is
compatible with the beta function for the gauge coupling to all orders \cite
{12-novikov}.

In order to say more about the values of the dynamical scales, in
particular the gaugino condensation scale $\Lambda_{N_c,0}$ in
 (\ref{dix-neuf}), we must know the parameters $m$, $b$, $\widetilde{b}$.
I give here a string-motivated example $m \sim m_{3/2}$, where $m_{3/2}$
is the gravitino mass and $(b,\widetilde{b})
\sim {1 \over M_P^{N_c-3}}$, where $M_P$ is the Planck mass. In this case,
in the $N_c \rightarrow \infty$
limit, for all the models discussed above, we get from (\ref{dix-neuf})
$\langle \lambda \lambda \rangle =
\Lambda_{N_c,0}^3 \sim m_{3/2} M_P^2$, as in the usual gaugino condensation
scenario in supergravity. Therefore, there is a hope that this way of
fixing dynamical scales is of phenomenological relevance.

Some remarks can be made concerning the duality proposed in \cite{4-seib435}%
, where it is argued that, for $\frac{3N_c}2<N_f<3N_c,$ the original
``electric'' theory with gauge group $SU(N_c)$ and $N_f$ flavours is dual to
a ``magnetic'' theory with gauge group $SU(N_f-N_c)$ and $N_f$ quark
flavours (similar dualities hold for $SO\left( N_c\right) $ or $SP\left(
N_c\right) $ gauge groups). The two different theories give the same physics
if there is a non-trivial $IR$ fixed point, where the two theories are
conformally invariant.

At this point, the parameters (and the fields) acquires anomalous dimensions
and we can define $m^{*}=\Lambda_{N_c,N_f}^{\frac{3N_c-N_f}{N_f}} \ m \ , \
 \left( b,%
\widetilde{b}\right) ^{*}=\Lambda_{N_c,N_f}^{\frac{\left( 3N_c-N_f\right)
N_c}{2N_f}%
}\left( b,\widetilde{b}\right) .$ Combining these fixed-point parameters
with the expression of the dynamical scale $\Lambda _{N_c,N_f}$ obtained
from (\ref{dix-sept}), we get

\begin{equation}
\prod\limits_{a=1}^{N_f-N_c}m_a^{*}= \left[ {\frac{\det
m_i^{*{\bar{\imath}}}}{%
\left( b \widetilde{b}\right)^{N_c \over 2}}}\right] ^{\frac 2{N_c-2}}\;.
\label{vingt-quatre}
\end{equation}

\noindent Notice that the scale $\Lambda $ cancelled out in (\ref
{vingt-quatre}), as it should in a conformally invariant theory.
Eq.(\ref{vingt-quatre})
defines an infrared surface, obtained as a result of considering $\Lambda_{N_c,
N_f}$ as a dynamical field.

In the magnetic theory, the symmetries and a correct decoupling force us to
introduce a new mass parameter $\mu_0 $ with charge $1$ under $U(1)_R,$ in
order to get a correct dynamical scale $\widetilde{\Lambda }%
_{_{N_f-N_c},_{N_f}}.$ The result is

\begin{equation}
\widetilde{\Lambda }_{_{N_f-N_c},_{N_f}}=\left( -1\right) ^{\frac{N_f-N_c}{%
2N_f-3N_c}}\mu_0 ^{\frac{N_f}{2N_f-3N_c}}\left( \prod\limits_am_a\right)
^{\frac 1{2N_f-3N_c}}\left[ \frac{\det m_i^{{\bar{\imath}}}}{{\left( b%
\widetilde{b}\right)}^{\frac{N_c}2}}\right] ^{-\frac 2{\left( N_c-2\right)
\left( 2N_f-3N_c\right) }}  \label{vingt-cinq}
\end{equation}

\noindent and combined with (\ref{dix-sept}) gives $\Lambda
_{N_c,N_f}^{3N_c-N_f}\;\widetilde{\Lambda }\;_{N_f-N_c,N_f}^{2N_f-3N_c}=%
\left( -1\right) ^{N_f-N_c}\mu_0 ^{N_f},$ in agreement with \cite{4-seib435}.
The scale $\mu_0 ,$ introduced for other reasons in \cite{4-seib435}, is
needed here in order to satisfy the decoupling theorem.

Hence we conclude that our main assumption, the gauge coupling to be a
dynamical chiral superfield, is compatible with the duality introduced in
\cite{4-seib435} and leads to the appearance of $IR$ surfaces of type (\ref
{vingt-quatre}).
\section{Effective Lagrangian approach for $N_f=N_c=N.$}

The symmetries of the supersymmetric $QCD$ for $N_f<N_c$ are
sufficient in order to exactly determine the low-energy effective
superpotential \cite{13-taylor}.\ For $N_f\geq N_c$ some additional,
dynamical information must supplement the symmetry behaviour.\ We will use
in the following some instanton arguments to reduce the freedom. The
Lagrangian (\ref{un}), (\ref{deux}) has a global symmetry only broken by the
$\left( b,\widetilde{b}\right) $ sources. It acts as $\left( \lambda ,Q,%
\widetilde{Q}\right) \rightarrow {\rm e}^{i\beta }\left( \lambda ,Q,%
\widetilde{Q}\right) ,\left( b,\widetilde{b}\right) \rightarrow {\rm e}%
^{(2-N_c)i\beta} \left( b,\widetilde{b}\right)  $ and leaves invariant
 the gauge fields, the fermions $\psi
_Q,\widetilde{\psi }_Q$ and the mass matrix $m_i^{{\bar{\imath}}}.$

The instanton effects then give rise in general to the formula

\begin{equation}
\left\langle {\det }M_{{\bar{\imath}}}^i-B\widetilde{B}%
\right\rangle =\Lambda _{N,N}^{2N}\sum_{n=0}^\infty a_n\frac{\left( b%
\widetilde{b}\right) ^{\frac{nN}{N-2.}}\Lambda _{N,N}^{2N}}{{\left( \det
m\right)}^{ \frac{2n}{N-2}}} \equiv \Lambda ^{2N}\cdot f\;,  \label{vingt-six}
\end{equation}

\noindent where $a_n$ is the coefficient of $N+1$ instantons contribution.
The formula (\ref{vingt-six}) is also compatible with all the symmetries (%
\ref{quatre}).\ In the $\left( b,\widetilde{b}\right) \rightarrow 0$ limit (%
\ref{vingt-six}) becomes the equation for the quantum moduli space $\left(
N_f=N_c\right) $ introduced in \cite{seib49}.\ The coefficient $a_0$ was
computed in the massless case to be $a_0=1$ \cite{7-finnell}.

Using the symmetries (\ref{quatre}) and the formula (\ref{vingt-six}), we
arrive at the following effective superpotential:

\begin{equation}
W_{{\rm eff}}=m_i^{\bar{\imath}}
M_{\bar{\imath}}^i+bB+\widetilde{b}\widetilde{B}+U\ln \frac{%
{\det }M_{{\bar{\imath}}}^i-B\widetilde{B}}{%
\Lambda _{N,N}^{2N}\cdot f(x)}\;,  \label{vingt-sept}
\end{equation}

\noindent

\noindent where $x=\frac{\left( b \widetilde{b}\right) ^{\frac N{N-2}}\Lambda
_{N,N}^{2N}}{{\left( \det m\right)}^{ \frac 2{N-2}}},$ and the field $S$ is
encoded in $\Lambda _{N,N}$ using (\ref{huit}).
Notice that by a change of variables we can move the function $f(x)$ from
the last term in (\ref{vingt-sept}) in the first two terms, such as to keep
valid the equation of the quantum moduli space introduced in \cite{seib49}.
Minimizing $W_{{\rm eff}}$
with respect to $S,M_{{\bar{\imath}}}^i,B,\widetilde{B}$ and U we find the
vacuum structure.

Defining by $x_0$ and $f_0$ the point and the value of the function $f$
at the minimum, we find the equations

\begin{eqnarray}
\Lambda _{N,N} &=&\left[ \frac{x_0\det m}{{\left( b\widetilde{b}\right)}^{\frac
N{2}}}\right] ^{\frac 1{N(N-2)}}\;, \ \left\langle U\right\rangle
=g^{-1}\left( \frac{\det m}{ b%
\widetilde{b} }\right) ^{\frac 1{N-2}}\;,  \label{vingt-huit} \\
\left\langle M_{{\bar{\imath}}}^i\right\rangle  &=&(-1)^{\frac 1{N-1}}\left[
x_0^{\frac 2{N-2}}f_0\cdot g\right] ^{\frac 1{N-1}}\left( \frac 1m\right) _{{%
\bar{\imath}}}^i\left( \frac{\det m}{ b\widetilde{b} }\right)
^{\frac 1{N-2}}\;,  \nonumber \\
\left\langle B\right\rangle  &=&g \ x_0^{\frac 2{N-2}} \ f_0\cdot \left(
\frac{\det m}{ b^{N-1}\widetilde{b} }\right) ^{\frac 1{N-2}}\;, \
\left\langle \widetilde{B}\right\rangle  =g \ x_0^{\frac 2{N-2}} \
f_0\cdot \left( \frac{\det m}{ b\widetilde{b}^{N-1} }\right)
^{^{\frac 1{N-2}}}\;.  \nonumber
\end{eqnarray}

\noindent $g$ is a function of $x_0$ and $f_0$ defined by the equation

\begin{equation}
(-1)^{\frac 1{N-1}}x_0^{\frac 2{\left( N-1\right) \left( N-2\right)
}}f_0^{\frac 1{N-1}}g^{\frac N{N-1}}+x_0^{\frac 2{N-2}}f_0 \ g^2=-1.
\label{vingt-neuf}
\end{equation}

The results (\ref{vingt-huit}), (\ref{vingt-neuf}) automatically fulfill the
Konishi anomaly (\ref{dix}).\ As we showed in the preceding paragraph, a
correct decoupling of massive quark flavours ask for $x_0=1,g=1.$ The
comparison of (\ref{vingt-huit}), (\ref{vingt-neuf}) with (\ref{sept}), (\ref
{onze}) allow the identification $C_B=f_0 \ , \ C_M=(-1)^{\frac 1{N-1}}
f_0^{\frac 1{N-1}}.$

Hence, what we need in order to explain in an effective Lagrangian approach
the results of the previous paragraph is the presence of non-linear terms
in the sources $m,b, \widetilde b$. This would be an exception of the
linearity principle postulated in \cite{5-intri50}. Nevertheless, the only
change in the results is in the stabilization of the dynamical scales, all
other dynamics being essentially the same.
Alternatively, we can redefine
 $\Lambda_{N,N}^{'}= {f(x)}^{1/2N} \Lambda_{N,N}$ by a non-linear change of
variables, in order to recover the usual effective Lagrangian of SQCD for
$N_f=N_c$ and all the known results \cite{seib49}, \cite{4-seib435},
\cite{15-kutasov}, \cite{16-intri}
(for fixed $\Lambda^{'}$). Then, the runaway behaviour would be just an
 artifact due to the singular value of the Jacobian of the
 change of variables.
\section{Conclusions}

The main goal of this letter is to apply methods based on holomorphy and
symmetries to the case where the gauge coupling constant is promoted to
a chiral superfield (the dilaton superfield). If its dynamics can be
described by an effective field theory, then its vacuum expectation value
can be exactly determined. In this way the dynamical scale of the theory
is exactly computed in $SQCD$ for $N_f \ge N_c$, $SQCD$ with an additional
chiral field $X$ in the adjoint of the gauge group $SU(N_c)$ and models with
$SO(N_c)$ gauge groups. Our main results (\ref{huit}),
(\ref{dix-sept}), (\ref{trente et un}), (\ref{trente-trois}) highly
constrain the dynamics and satisfy non-trivial consistency checks, for
example (\ref{neuf}), (\ref{douze}), (\ref{dix-huit}), (\ref{vingt-deux})
and (\ref{vingt-quatre}). For $N_f < N_c$ we find the usual runaway
behaviour.

The actual value of the gaugino condensation scale (\ref{dix-neuf}) is
shown to be potentially of phenomenological interest in the limit
$N_c \rightarrow \infty$, for hidden sector models of supersymmetry
breaking in supergravity.

In an effective Lagrangian approach, these results can be recovered if
there are non-linear terms in the chiral background sources (\ref{deux}),
(\ref{treize}), (\ref{trente}) and (\ref{trente-deux}).
Usually, these non-linear terms are eliminated in the literature by a
redefinition of the dynamical scale, which is perfectly justified in
$N=1$ supersymmetric theories with a fixed gauge coupling constant
(for a discussion on this point,
see \cite{17-seib}). We argued here that if the gauge coupling is
considered as a dynamical field, we must keep the non-linear
structure and the picture which emerges is consistent.

We hope the results of this letter to be relevant for the dilaton
stabilization problem of the effective string theories.

\vskip 1.2cm
{\bf Acknowledgements}
\vskip 0.3cm
It is a pleasure to thank P. Bin\'etruy, M. Chemtob, J. Louis, R. Peshansky,
C. Savoy and A. Vainshtein for helpful discussions and comments.

\end{document}